\documentstyle[12pt,aps,prd,preprint,psfig,floats]{revtex}
\begin{document}
\draft
\title{The evolution of density perturbations in two quintessence models}
\author{Tame Gonzalez\thanks{tame@uclv.edu.cu} and Rolando Cardenas\thanks{rcardenas@uclv.edu.cu}}
\address{Physics Department. Las Villas Central University. Santa Clara. Cuba}
\maketitle
\begin{abstract}
In this work we investigate the evolution of matter density
perturbations for two different quintessence models. One of them
is based on the Einstein theory of gravity, while the other is
based on the Brans-Dicke scalar tensor theory. We show that it is
possible to constrain the parameter space of the models using the
determinations for the growth rate of perturbations derived from
data of the 2-degree Field Galaxy Redshift Survey.
\end{abstract}
\section{Introduction}
In the past few years, it has become apparent that the energy
budget of our universe is dominated by an unknown component called
"dark energy". The WMAP table of "best" cosmological parameters
\cite{lambda}, for instance, gives a $0.73\pm0.04$ abundance for
it, and a value of its equation of state $\omega<-0.78$. In order
to relieve some problems of the popular $\Lambda CDM$ model (like
the fine tuning issue), a dynamical $\Lambda$-term has been
proposed as representative of the dark energy. It's more popular
version is a slowly rolling scalar field named quintessence.
Many alternative cosmological models have been proposed, and
indeed it is a challenge the work of ruling out all the "incorrect
ones" on observational grounds. For instance many different
potentials for these self interacting scalar fields (quintessence)
have been proposed. However, it is obvious the importance of the
observational exploration of the proposed cosmological models, and
in this paper we give a further step in this direction.
In \cite{Isra1} two cosmological models are proposed. In both
cases two fluids fill the universe: a background fluid of ordinary
matter and a self-interacting scalar-field fluid accounting for
the dark energy, both in Einstein's theory and in Brans-Dicke
gravity. A linear relationship between the Hubble expansion
parameter and the time derivative of the scalar field is used to
derive exact cosmological attractor-like solutions. And a priori
assumptions about the functional form of the potential or the
scale factor behavior are not necessary. All these features render
two very interesting models that motivated us to proceed with the
observational check of them.
In the original paper, the authors made a first check with some
observational facts, related with Cosmic Microwave Background
(CMB), nucleosynthesis, structure formation (restriction on
quintessence density during galaxy formation epoch) and type Ia
supernovae (SN Ia). Now we proceed the observational checking
considering another aspect of structure formation: the galaxy
motions and clustering, i. e., the evolution of density
perturbations in the Universe.
In the past few years, observations of the large scale structure
of the Universe have improved greatly. The development of
fiber-fed spectrographs that can simultaneously measure spectra of
hundreds of galaxies has provided large redshift surveys such as
the 2-degree Field Galaxy Redshift Survey (2dFGRS) and the Sloan
Digital Sky Survey (SDSS).
In particular, the Anglo-Autralian Telescope of the 2dFGRS has
obtained the redshift of a quarter million galaxies. This
collaboration has produced abundant data and technical papers
\cite{2dFGRS} about galaxy motions and clustering, and we will
refer to some of this, in particular their velocity/density
comparisons.

The paper is organised as follows: in section II we outline the
main characteristics of the models, in section III the main
aspects of velocity/density comparisons are exposed and the
equation for the growth of perturbations is solved, in section IV
the observational check is presented and interpreted, while in
section V conclusions are drawn. Finally, references are supplied.

\section{The Models}
In this section we supply the main equations characterizing the
dynamics of the models. For a detailed description of them, we
refer to \cite{Isra1}. As said above, one of the models is based
on the Einstein gravity and the other in the Brans-Dicke one.
\subsection{Einstein Theory}
We analyze the solution in Einstein's theory with two fluids: a
background fluid of ordinary matter and a self-interacting scalar
fluid that accounts for the dark energy component in the universe
that Arias et. all. studied in \cite{Isra1}.
In this case the ansatz relating the expansion parameter and the
time derivative of the scalar field is:
\begin{equation}
H=k\dot{\phi},
\end{equation}
where $k$ is a constant parameter.
 The exact solution for,
respectively, the scale factor, the Hubble expansion parameter,
the matter density and equation of state are
\begin{equation}
a(\tau)=\{\sqrt{\frac{A}{B}}sinh[\mu(\tau+\tau_{0})]\}^{\frac{2k^{2}}{3k^{2}\gamma-1}}
\end{equation}
\begin{equation}
H(\tau)=\sqrt{\frac{8k^{2}}{6k^{2}-1}\xi_{0}}a(\tau)^{-1/2k^{2}}coth[\mu(\tau+\tau_{0})]
\end{equation}
\begin{equation}
\Omega_{m}(\tau)=(1-\varepsilon)\{cosh[\mu(\tau+\tau_{0})]\}^{-2}
\end{equation}
\begin{equation}
\omega_{\phi}(\tau)=-1+\frac{1}{3k^{2}(1-\Omega_{m}(\tau))}
\end{equation}
whit $dt=a^{\frac{1}{2k^{2}}}d\tau$.
This cosmological model depends only on three free parameters
($\gamma, k, \varepsilon$). The intermediate parameters $A, B,
\mu, \tau_0$ and $\xi_0$ are determined from the normalisation
used or recast into ($\gamma, k, \varepsilon$). $\gamma$ is the
barotropic index of the matter fluid. In this paper we study the
evolution of matter density perturbations, well into the epoch of
matter domination over radiation, so we set $\gamma=1$. As seen
from (1), $k$ is the parameter relating the expansion rate and the
time derivative of the scalar field , and $\varepsilon$ is the
density $\Omega_{\phi}(z)$ of dark energy in the early stages of
the evolution ($z\sim\infty$) provided the universe is flat. From
CMB, nucleosynthesis and galaxy formation observations, in
\cite{Isra1} the authors additionally constrained this parameter
space to be ($\gamma=1, k>0, 0\leq\varepsilon\leq0.045$).
Interesting to note that SN Ia was not useful to constrain the
parameter space. Indeed, it is well known the controversy about
the degeneracy of supernovae observations
\cite{w,mb,mb2,w2,w3,mb3}.
At this model we will call Model E (Einstein gravity).
\subsection{Brans-Dicke gravity}
In this model we analyse the exact solution in gravity with
non-minimal coupling between the quintessence field and the
background fluid, in particular the Brans-Dicke theory with two
fluids: a barotropic perfect fluid of ordinary matter and a
self-interacting scalar fluid accounting for the dark energy in
the universe.
In this case we are faced with two relevant frames (the Jordan
frame and the Einstein frame), in which Brans-Dicke theory can be
formulated. There has been discussion on whether these two frames
are equivalent \cite{equivalen}. It is not our aim to participate
in this controversy and for practical reasons (simplicity of
mathematical handling) we chose the Einstein frame.
Now the ansatz relating the expansion parameter and the time
derivative of the (rescaled) scalar field is:
\begin{equation}
\bar{H}=\frac{\dot{\psi}}{\lambda},
\end{equation}
where $\lambda$ is a constant parameter. The bar notation means we
are working in the Einstein frame of BD theory.
In this frame, the exact solutions for, respectively, the scale
factor, the matter density and equation of state are
\begin{equation}
\bar{a}(r)=\{\sqrt{\frac{A}{B}}sinh[\mu(r+r_{0})]\}^{\frac{2}{\lambda(\delta-\lambda)}}
\end{equation}
\begin{equation}
\bar{\Omega}_{m}(r)=n^{2}(1-\varepsilon)\{cosh[\mu(r+r_{0})]\}^{-2}
\end{equation}
\begin{equation}
\bar{\omega}_{\psi}(r)=-1+\frac{\lambda^{2}}{3(1-\bar{\Omega}_{m}(r))}
\end{equation}
where $d\bar{t}=\bar{a}^{-\frac{\lambda^{2}}{2}}dr$.
This cosmological model also depends on three free parameters
($\gamma, \lambda, \varepsilon$). Like in the former case, the
intermediate parameters $A, B, \mu, r_0, \delta$ and $n$ are
determined from the normalisation used or expressed through
($\gamma, \lambda, \varepsilon$). Also, from CMB, nucleosynthesis
and galaxy formation observations, in \cite{Isra1} the authors
constrained this parameter space to be ($\gamma=1, 0<\lambda<0.37,
0\leq\varepsilon\leq0.045$). Again SN Ia observations were not
useful to constrain the parameter space.
To this case we will call Model BD (Brans-Dicke gravity).
\section{Perturbation growth}
We will study the evolution of mass density contrast
$(\delta=\delta\rho/\rho)$ in the mass distribution, modelled as a
pressureless fluid, in linear perturbation theory. This method is
based on Newtonian mechanics, that is better suited to the study
of the development of structure such as galaxies and clusters of
galaxies. This computation requires that we be able to isolate a
region small enough for the Newtonian gravitational potential
energy and the relatives velocities within the region to be small
(non relativistic) \cite{peebles}.
The equation of time evolution of mass density contrast is
\begin{equation}
\ddot{\delta_{m}}+2H\dot{\delta_{m}}-4 \pi G \rho_{m} \delta_{m}=0
\end{equation}
Where the dot means derivative with respect to the comoving time.
In this equation the relative contribution of dark energy to the
energy budget enters into the expansion rate $H$. We shall
consider this equation in the matter dominated era, when the
radiation contribution is really negligible.
The linear theory relates the peculiar velocity field v and the
density contrast by
\begin{equation}
v(x)=H_{0} \frac{f}{4\pi} \int \delta_{m}(y) \frac{x-y}{|x-y|^3}
d^3y
\end{equation}
where the growth index $f$ is defined as
\begin{equation}
f\equiv \frac{d\ln\delta_{m}}{d\ln a}
\end{equation}
where $a$ is the scale factor.
To solve the equation (10) for the evolution or perturbations we
introduced several changes of variable, allowing to rewrite it in
the form
\begin{equation}
X^{2}(1+X^{2})\ddot{\delta_{m}}+X
\dot{\delta_{m}}(X^{2}d+e)-c\delta_{m}=0
\end{equation}
where $X=sinh[\mu(\tau+\tau_{0})]$, the dot means derivative with
respect to X, and $c$, $d$ and $e$ are constants characteristic of
each model.
Equation (13) has two linearly independent solutions, the growing
mode $\delta_{m+}$ and the decreasing mode $\delta_{m-}$, which
can be expressed in terms of hypergeometric functions of second
type $ _{2}F_{1}$. We get
\begin{equation}
\delta_{m+} \propto X^{1/2\{\sqrt{A}+1-e\}}
\cdot_{2}F_{1}[\frac{1}{4}-\frac{e}{4}+\frac{1}{4}\sqrt{A},-\frac{1}{4}+\frac{d}{2}-\frac{e}{4}+\frac{1}{4}\sqrt{A},1+\frac{1}{2}\sqrt{A},-X^{2}]
\end{equation}
\begin{equation}
\delta_{m-} \propto X^{1/2\{-\sqrt{A}+1-e\}}
\cdot_{2}F_{1}[\frac{1}{4}-\frac{e}{4}-\frac{1}{4}\sqrt{A},-\frac{1}{4}+\frac{d}{2}-\frac{e}{4}-\frac{1}{4}\sqrt{A},1-\frac{1}{2}\sqrt{A},-X^{2}]
\end{equation}
where $A=(e-1)^{2}+4c$ and .
For $\tau \ll 1$ we can write
\begin{equation}
\delta_{m+} \propto X^{1/2\{1+\sqrt{A}-e\}},\delta_{m-} \propto
X^{1/2\{1-\sqrt{A}-e\}}
\end{equation}
For determining the growth index of the perturbations we use the
growing mode $\delta_{m+}$ (equation (16)) and substitute into
equation (12).
It is well known the biasing effect in galaxy formation, i. e.,
the relative perturbations in the galaxy field and the matter
field, on a point-by-point basis, are not equal:
\begin{equation}
\frac{\delta n}{n}(x)=b\frac{\delta \rho}{\rho}(x)
\end{equation}
where $n$ refers to the galaxy number density and $b$ is the bias
parameter.
The parameter $\beta=\frac{f}{b}$ relates the growth rate $f$ of
the perturbations (and hence the velocity field of the galaxy
motions) with the density bias $b$. In this sense astrophysicists
speak on a velocity/density comparison. Indeed, a compelling
agreement is seen to exist between the velocity and density
fields, which offers one possible test for the gravitational
instability picture for the origin of structure \cite{ll}.

The 2dFGRS has obtained redshift of a quarter million galaxies
with an effective depth of $z=0.17$. From a precise measurement of
the clustering, a value of $\beta=0.43\pm 0.07$ has been
determined \cite{peacock}. The bias parameter has been measured by
computing the bispectrum of the 2dFGRS, it is $b=1.04\pm 0.11$
\cite{verde}. Combining the measurements of $\beta$ and $b$, the
growth index at $z=0.17$ can be estimated: it is
$f(z=0.17)=0.45\pm 0.06$ \cite{verde}.
The intermediate parameters in equation (16) for the growing mode
of density perturbations can be recast into the parameter spaces
of our models ($\gamma, k, \varepsilon$) and ($\gamma, \lambda,
\varepsilon$), respectively. After using equation (12), this
results that the growth rate $f$ will have the same parametric
dependence. Now we use this fact to additionally constrain these
spaces.

\section{Observational Check}
\subsection{Model E}
As said above, in the original paper ($\gamma, k, \varepsilon$)
had been already constrained to ($\gamma=1, k>0,
0\leq\varepsilon\leq0.045$). Applying equation (12) does not give
further constrain on $\varepsilon$, but it does on $k$, as seen in
Figure 1.
\begin{figure}[tbh!]
\centerline{ \psfig{figure=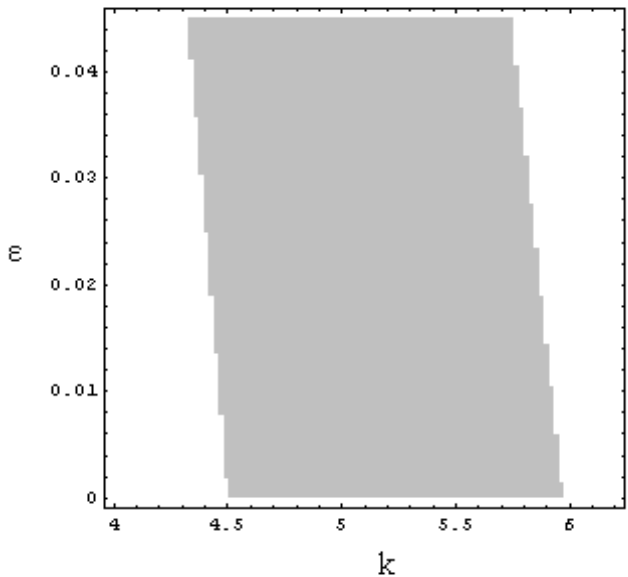,width=0.5\textwidth,angle=0}
}
\bigskip
\caption{Parameter Space for the model based on Einstein gravity.
The parameter $k$ is now constrained to a rather narrow region}
\label{fig1:hubble}
\end{figure}
Assuming a flat universe, the parameter
$\varepsilon=\Omega_\varphi(\infty)$, is the amount of dark energy
in the very early universe ($z\sim\infty$). It is known that an
appreciable amount of dark energy at that epoch would imply an
expansion fast enough to prevent the formation of structure at
$z\sim3$, but this parameter had been already constrained in the
original paper, so it is not problematic that now it has not been
additionally constrained. On the other hand, as showed in
\cite{Isra1}, the condition (1) acts like an interesting selection
principle that, amongst the possible critical points, preserves
only the attractor-like solutions. In this case, the
velocity/density comparison allows to locate $k$ in a rather
narrow region, thus acting as a stringent selector of
attractor-like solutions. Two things are to be said. Firstly,
Figures 1 shows the joint region ($k, \varepsilon$) satisfying the
velocity/density comparison, for $k$ alone we might have a wider
variation. Secondly, though narrow the interval, still we can
speak of $k$ as a selector of a class of solutions. Of course,
with this models the authors don't pretend to definitely solve the
fine tuning issue arising in cosmology.

\subsection{Model BD}
 In the figure 2 we show the parameter space that we
obtained for the model based on Brans-Dicke gravity. In this
figure the region inside straight line is the parameter space
obtained in \cite{Isra1}, while the one inside the dashed line is
the parameter space obtained considering the perturbation  growth.
Then the shaded region gives the new parameter space, obeying all
constrains considered so far.
\begin{figure}[tbh!]
\centerline{ \psfig{figure=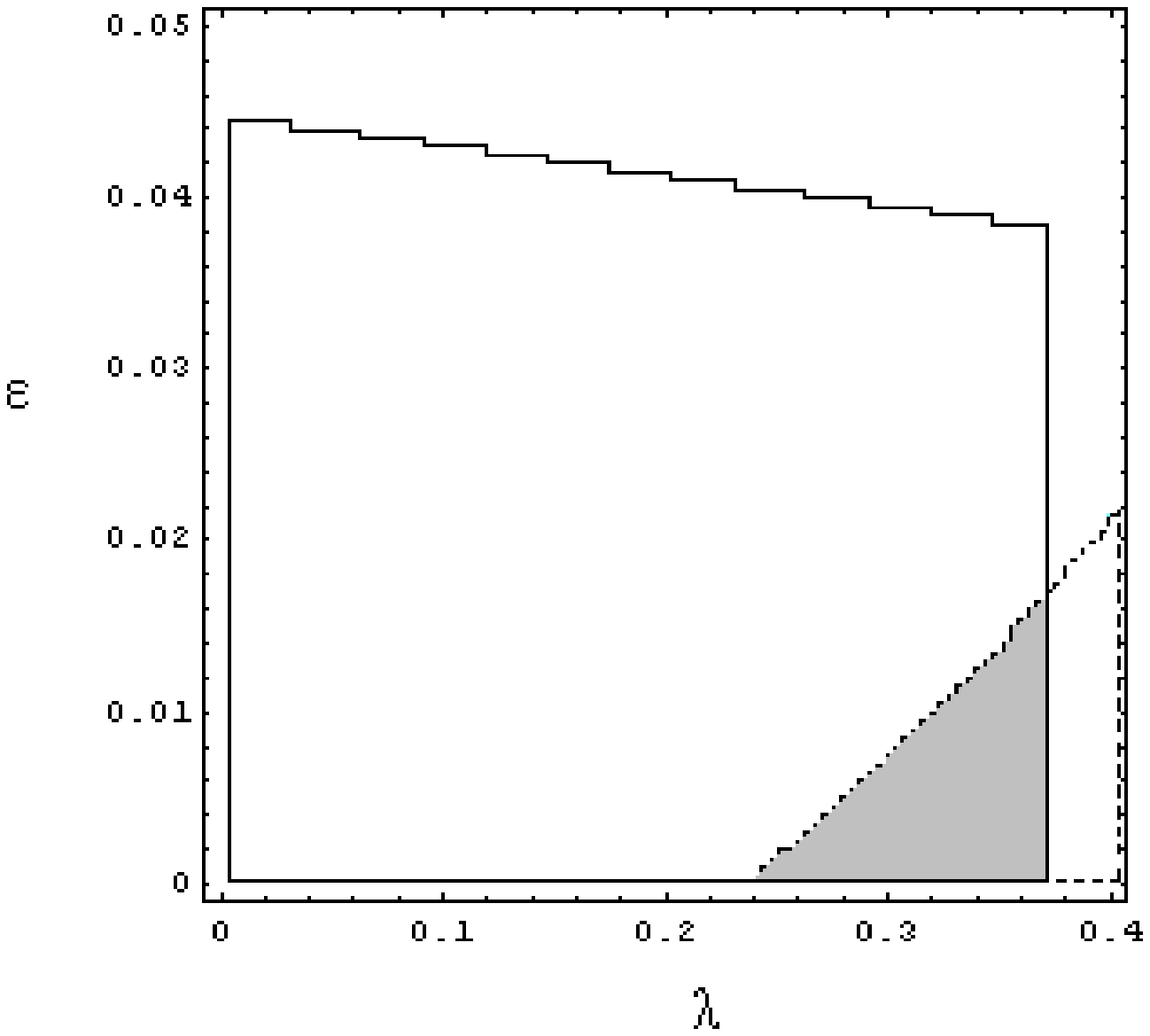,width=0.5\textwidth,angle=0}
}
\bigskip
\caption{Parameter Space for BD gravity. The region inside the
straight line marks original constrains, while the region inside
the dashed line signals the constrain from perturbations growth}
\label{fig1:hubble}
\end{figure}
The parameters ($\lambda, \varepsilon$) of this model have a
similar physical interpretation of ($k, \epsilon$).
In this case, both parameters were already considerably
constrained from the original observations. The consideration of
the growth rate of perturbations now gives additional
restrictions, as seen from Figure 2. The above interpretation of
$k$ as a stringent selector of a class of attractor-like solutions
also holds for $\lambda$.
\section{Conclusions}
In the models presented here we found that the study of
perturbation growth is a good tool to constraint the parameter
space. Research on the origins and evolution of the large-scale of
the universe is one of the hottest topics in cosmology. In this
work, we have used the relation between the peculiar velocity
field of the galaxies, the growth rate of perturbations and the
density bias in galaxy formation to make another step in the
observational check of two quintessence models, resulting in a
further constrain on the parameter space of the models.
We plan in the oncoming future proceed this works using another
cosmological probes, like CMB, for instance.

\end{document}